\documentclass[aps, prl, twocolumn, superscriptaddress, longbibliography]{revtex4-2}
\usepackage[colorlinks, linkcolor=blue, anchorcolor=blue, citecolor=blue]{hyperref}
\usepackage{amsmath}
\usepackage{amssymb}
\usepackage{graphicx}
\usepackage{braket}
\usepackage{multirow}
\usepackage{color}
\usepackage{soul}
\usepackage{diagbox}
\usepackage{rotating}
\usepackage{gensymb}
\usepackage{makecell}

\usepackage{ulem}%only for the command \sout = scrap
\begin{document}

\title{Global magnetic phase diagram and multiple unconventional magnets in NiAs-type compounds}
\author{Shibo Shen}
\affiliation{School of Emerging Technology, University of Science and Technology of China, Hefei 230026, China}
\affiliation{Hefei National Laboratory, Hefei 230088, China}
\affiliation{New Cornerstone Science Laboratory, University of Science and Technology of China, Hefei, 230026, China}

\author{Yilin Wang}\email{yilinwang@ustc.edu.cn}
\affiliation{School of Emerging Technology, University of Science and Technology of China, Hefei 230026, China}
\affiliation{Hefei National Laboratory, Hefei 230088, China}
\affiliation{New Cornerstone Science Laboratory, University of Science and Technology of China, Hefei, 230026, China}

\date{\today}

\begin{abstract}
NiAs-type compounds such as CrSb and MnTe host $g$-wave altermagnet (AM) state. In order to search other possible unconventional magnets in this system, we present a global magnetic phase diagram based on a classical $J_1$-$J_2$-$J_3$ Heisenberg model and density functional theory (DFT) calculations. We find another $g$-wave AM state and two $f$-wave OPMs in the phase diagram. Intriguingly, we show that a mixed-parity of the $f$-wave OPM and $g$-wave AM state can naturally emerge in an umbrella-like noncollinear magnetic structure. Our DFT calculations show that CrSe and CrTe$_{1-x}$Se$_x$ are generally in such mixing state with dominated $f$-wave component. The interlayer next-nearest-neighbor coupling $J_3$ is shown to be crucial in determining the phase diagram and in inducing strong competition between conventional and unconventional magnets. Inspired by this, we demonstrate that AM or OPM could be realized by applying chemical doping or strain to conventional magnets. Our results provide a guidance for design of both even- and odd-parity as well as mixed-parity unconventional magnets in NiAs-type compounds.
\end{abstract} 

\maketitle

\textit{Introduction.} The discovery of altermagnet (AM) has drawn extensively attention to the unconventional antiferromagnets (AFM) with non-relativistic spin-splitting in momentum space but without net magnetization~\cite{MnO2_DFT,2019_THEORY,V2Se2O_theory,theory_RbVTeO,theory_prx_1,theory_prx_2,Anomalous_Hall,review_am,largescale_DFT,review_yao,Landa_Theory_of_Altermagnetism,theory_AM,Magnetic_geometry}, due to potential applications, such as spintronics~\cite{Electricalspinsplitter,GMR_theory,GateFieldControl,180switchingneelvector,altermagneticorder,controlSST}, multifunctional materials~\cite{ZhouTong:2025,LiuQihang:2025,MaYanming:2025}, and unconventional superconductivity~\cite{NSM,TSC_1,FFLO,FFLO_3,TSC_2,FFLO_2,pwavesc,FFLO_4,TSC_3}. The spin space group is the tool for describing such states~\cite{spingroup1,spingroup2,spingroup3,spingroup4,spingroup5,spingroup6,spingroup_lqh,luo2026spingroup_1,luo2026spingroup_2}. According to the parity of the spin textures in momentum space, they are generally classified into even-parity [$\mathbf{s}(\mathbf{k})=\mathbf{s}(\mathbf{-k}$), $d,g,i$-wave] and odd-parity [$\mathbf{s}(\mathbf{k})=-\mathbf{s}(\mathbf{-k}$), $p,f,h$-wave]. Altermagnet belongs to even-parity with broken time reversal $T$ but preserved spatial inversion $P$ in band structures,   and its zero magnetization is guaranteed by real-space rotational transformation rather than by translation or inversion. Many altermagnet candidates have been proposed, such as $d$-wave RuO$_2$~\cite{ruo2_theory,ruo2_anomalous_Hall,Transport_ruo2,ruo_arpes_2,ruo_arpes_1,ruo2_Magnetoresistance,ruo2_Transport}, $A$V$_2$Se$_2$O family~\cite{theory_RbVTeO,KV2Se2O_ARPES,RbVTeO_arpes,CsVTeO,KVSeO_stm,KVSeO_stm2,CsV2Se2O_stm,Caoguanghan:2026}, $g$-wave CrSb~\cite{CrSb_arpes_3,CrSb_arpes_1,CrSb_arpes_2,CrSb_arpes_4,CrSb_arpes5} and MnTe~\cite{MnTe_arpes_3,MnTe_arpes_1,MnTe_arpes_2,MnTe_arpes_4,MnTe_arpes5,PPEM_MnTe}, just to name a few. Odd-parity magnets (OPM), in contrast, with preserved $T$ but broken $P$ in band structures, are also proposed~\cite{p_wave_magnets,Jacob:2024,Mitscherling2026}. They exhibit various properties such as transport anisotropy~\cite{p_wave_magnets,Motohiko:2025}, Edelstein effect~\cite{Chakraborty2025Highly,JairoEdelstein:2025} and unconventional superconductivity~\cite{Jacob:2025,Fukaya_2025,Khodas:2026,KTLaw:2025,Yukio:2024,Kim2026-nv,Pal2026-ed}. Materials such as CeNiAsO~\cite{p_wave_magnets,Mitscherling2026,nonsymmorphic}, Mn$_3$GaN~\cite{p_wave_magnets}, NiI$_2$~\cite{NiI2}, Gd$_3$(Ru$_{1-\delta}$Rh$_{\delta}$)$_4$Al$_{12}$~\cite{GdRuRhAl}, Fe-based superconductors~\cite{nonsymmorphic,Morten:2026} have been predicted. However, the direct observation of spin splitting in these OPMs by experiments such as ARPES have not been reported, probably due to the complexity of these materials. Thus, more OPM candidates, especially those with quasi-two-dimensional and metallic nature, need to be designed theoretically.

The widely studied $g$-wave AM CrSb and MnTe belong to the family of NiAs-type compounds. This system features nonsymmorphic space group (P6$_3$/mmc) and triangular magnetic sublattice. The inversion operator is usually nontrivial in the nonsymmorphic space group~\cite{nonsymmorphic} and the frustration of triangular lattice usually results in noncollinear (NCL) magnetic structure which are both crucial for realizing OPM~\cite{luo2026spingroup_1,Triangular_Lattice2,f_wave,TSC_3,Triangular_Lattice}. This indicates that the NiAs-type compounds should be also a good platform for hosting OPMs. However, a global magnetic phase diagram guiding the search and design of these unconventional magnets in this system is not yet available. A more interesting question is whether a mixed-parity of the AM and OPM magnets could naturally realize or not in this system. 

In this paper, in order to search other possible unconventional magnets, in particular, OPMs in NiAs-type compounds, we present a global magnetic phase diagram based on a  classical $J_1$-$J_2$-$J_3$ Heisenberg model and density functional theory (DFT) calculations. We find two $g$-wave AMs and two $f$-wave OPMs in this system. Intriguingly, we demonstrate that an umbrella-like NCL magnetic structure~\cite{CrSeneutron,CrSe_oddparity} is actually a parity-mixing of $f$-wave OPM and $g$-wave AM state. Our DFT calculations show that CrSe and CrTe$_{1-x}$Se$_x$ are generally in such mixing state with dominated $f$-wave component. We show the important roles of the interlayer next-nearest-neighbor coupling $J_3$ in inducing strong competition between conventional and unconventional magnets. Inspired by this, we show that AM and OPM could be realized by applying chemical doping or strain to conventional FM or AFM states. We thus provide a feasible route for search and design of both even- and odd-parity as well as mixed-parity unconventional magnets in the NiAs-type compounds.

 \begin{figure*}
    \centering
    \includegraphics[width=1.0\textwidth]{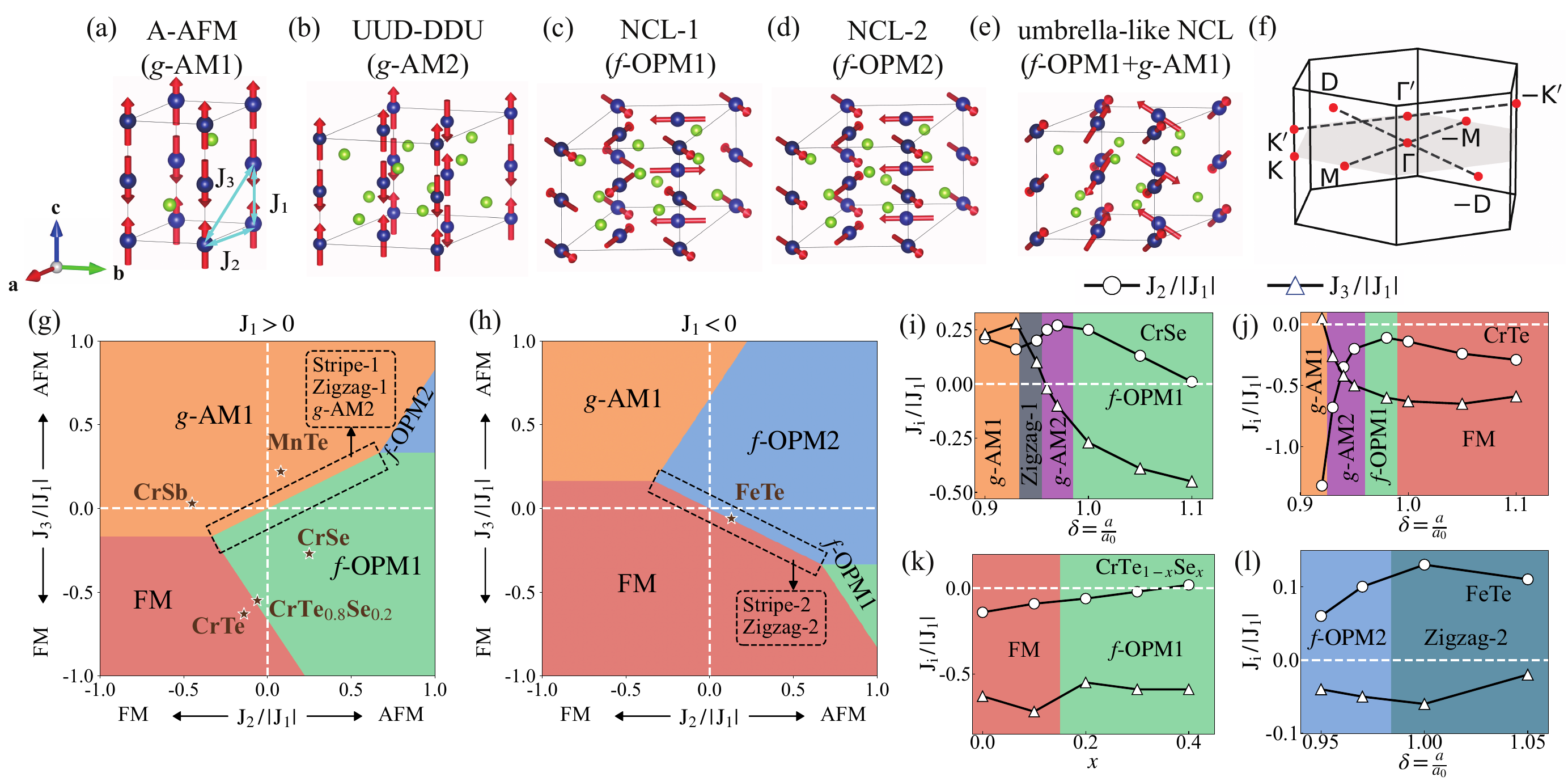}
    \caption{
     (a)-(e) Five unconventional magnetic structures. (f) Brillouin zone and $k$-path for plotting band structures. (g) and (h) $J_2/|J_1|-J_3/|J_1|$ phase diagrams of the $J_1$-$J_2$-$J_3$ model for AFM and FM coupling of $J_1$, respectively. (i)-(l) DFT calculated $J_2/|J_1|$, $J_3/|J_1|$ and ground state of CrSe, CrTe, CrTe$_{1-x}$Se$_x$ and FeTe as functions of lattice parameters $a$ or chemical compositions. $J_1>0$ in CrSe, CrTe and CrTe$_{1-x}$Se$_x$, while $J_1<0$ in FeTe. Note that CrSe and CrTe$_{0.8}$Se$_{0.2}$ are generally in the umbrella-like mixed-parity state, with a very small component of $g$-AM1 in addition to the dominated $f$-OPM1 component.
     }
    \label{fig:phase}
\end{figure*}
\textit{Phase diagram of classical $J_1$-$J_2$-$J_3$ model.} We begin the analysis of the possible magnetic states in NiAs-type compounds based on a minimal classical $J_1$-$J_2$-$J_3$ Heisenberg model. As shown in Fig.~\ref{fig:phase}(a), $J_1$ and $J_3$ are the interlayer nearest-neighbors (NN) and next-nearest-neighbors (NNN) couplings, respectively, and $J_2$ is the intralayer NN coupling. We use the convention that $J_{i}>0$ ($J_{i}<0$) for AFM (FM) coupling. Besides FM state, five unconventional AFM states could exist in the $1\times1\times1$ or $\sqrt{3}\times\sqrt{3}\times1$ supercell. As shown in Fig.~\ref{fig:phase}, they are (a) collinear A-type AFM; (b) UUD-DDU, intralayer collinear AFM order with two spin up and one spin down, and with interlayer AFM order; (c) NCL-1, coplanar $120{\degree}$-noncollinear AFM order and interlayer AFM order, (d) NCL-2, analogous to NCL-1 but with interlayer FM order. Generally, canted  moments along the $c$-axis in NCL-1 could be possible, which leads to a noncoplanar umbrella-like NCL state, as shown in Fig.~\ref{fig:phase}(e). We also consider conventional collinear stripe and zigzag AFM states that may also occur in triangular lattices~\cite{YongsongLuo:2025}, in $2\times1\times1$ and $2\times3\times1$ supercells (we term Stripe-1/Zigzag-1 and Stripe-2/Zigzag-2 for interlayer AFM and FM orders, respectively. See Supplementary Materials~\cite{suppl}).

The A-type AFM is the extensively studied $g$-wave altermagnet in CrSb and MnTe, and the UUD-DDU is also a $g$-wave altermagnet identified by the online tool \textit{FINDSPINGROUP}~\cite{FINDSPINGROUP}, so we label them as $g$-AM1 and $g$-AM2, respectively. We find that the spin space group of NCL-1 and NCL-2 do not contain the combined symmetry $PT$ or multiple spin-rotation-translation symmetries with distinct rotation axes, but contain symmetry operations like $\{U_z (\theta)||I|\tau\}$ and $\{TU_z(\pi)||I\}$, where $U_z(\theta)$ is spin rotation around $z$-axis, $\tau$ is a spatial fractional translation, and $I$ is spatial identity. According to the symmetry criteria~\cite{luo2026spingroup_1,luo2026spingroup_2}, NCL-1 and NCL-2 are OPMs with a $f$-wave spin-texture ($s_z(\mathbf{k})=-s_z(\mathbf{-k}), s_{x,y}\mathbf(\mathbf{k})=0$). We label them as $f$-OPM1 and $f$-OPM2, respectively. In the umbrella-like NCL structure, the symmetry $\{TU_z(\pi)||I\}$ is broken and there are no other symmetries satisfying the condition for OPM due to the additional out-of-plane moments. However, these out-of-plane components actually form a A-AFM structure ($g$-AM1 state), so this umbrella-like state is a parity-mixing of $f$-OPM1 and $g$-AM1 states. Since both states have nonzero spin-texture of only $s_z(\mathbf{k})$, its spin-texture is a mixing of $f$- and $g$-waves of $s_z(\mathbf{k})$. 

\begin{table*}
    \centering
    \caption{The DFT calculated magnetic exchange couplings ($|\vec{S}_i|=1$, meV) and total energies (per the $1\times 1\times 1$ cell, meV) of nine magnetic states, for several NiAs-type materials. The numbers in the parenthesis are the total energies calculated by Eqns.~\ref{eqn:1}-\ref{eqn:6} with the calculated $J_i$ as inputs. }
    \begin{ruledtabular}
    \begin{tabular}{ccccccccccccc}

    Material & $J_1$ & $J_2$ & $J_3$ & $g$-AM1 & $g$-AM2 & $f$-OPM1 & $f$-OPM2 & FM & Stripe-1 & Stripe-2 & Zigzag-1 & Zigzag-2\\
         \hline
    CrSb & 74.9 & -34.0 & 2.2 & \textbf{\makecell{0(0)}} & \makecell{325 (307)} & \makecell{350 (346)} & \makecell{635 (619)} & \makecell{350 (352)} & \makecell{307 (307)} & \makecell{588 (589)} & \makecell{313 (307)} & \makecell{578 (589)}\\

    MnTe & 42.0 & 3.2 & 9.2 & \textbf{\makecell{0{(0)}}} & \makecell{132 (122)} & \makecell{154 (137)} & \makecell{213 (194)} & \makecell{387 (389)} & \makecell{120 (122)} & \makecell{216 (216)} & \makecell{118 (122)} & \makecell{212 (216)}\\

    CrSe & 30.3 & 7.7 & -8.1 &\makecell{250{(215)}} & \makecell{18 (24)} & \textbf{\makecell{0 (0)}} & \makecell{256 (218)} & \makecell{177 (142)} & \makecell{58 (24)} & \makecell{239 (210)} & \makecell{30 (24)} & \makecell{235 (210)}\\

    CrTe$_{0.8}$Se$_{0.2}$ & 27.0 & -1.7 & -14.9 & \makecell{251{(253)}} & \makecell{18 (28)} & \textbf{\makecell{0 (0)}} & \makecell{232 (287)} & \makecell{6 (3)} & \makecell{39 (28)} & \makecell{213 (255)} & \makecell{26 (28)} & \makecell{199 (255)}\\

    CrTe & 21.0 & -3.0 & -13.2 &  \makecell{189{(233)}} & \makecell{18 (46)} & \makecell{7 (22)} & \makecell{197 (265)} & \textbf{\makecell{0 (0)}} & \makecell{31 (46)} & \makecell{188 (235)} & \makecell{25 (46)} & \makecell{178 (235)}\\

    FeTe & -95.4 & 12.3 & -5.4 &  \makecell{358{(525)}} & \makecell{198 (340)} & \makecell{145 (317)} & \makecell{3 \textbf{(0)}} & \makecell{36 (14)} & \makecell{280 (340)} & \makecell{25 (2)} & \makecell{228 (340)} & \makecell{\textbf{0} (2)}\\

    FeTe(0.97a$_0$) & -74.8 & 7.3 & -4.1 &\makecell{415{(398)}} & \makecell{244 (274)} & \makecell{186 (258)} & \makecell{\textbf{0} (8)} & \makecell{17 \textbf{(0)}} & \makecell{291 (274)} & \makecell{24 (7)} & \makecell{280 (274)} & \makecell{5 (7)}
    \end{tabular}
\end{ruledtabular}
    \label{tab:energy}
\end{table*}

The total energies derived from this classical $J_1$-$J_2$-$J_3$ model are (per $1\times1\times1$ cell, assuming $|\vec{S}_i|=1$),
\begin{eqnarray}
E_{g\text{-AM1}}&=&-2J_1 + 6J_2 -12J_3,\label{eqn:1}\\
E_{g\text{-AM2}/\text{Stripe-1}/\text{Zigzag-1}}&=&-2J_1 - 2J_2 +4J_3,\label{eqn:2}\\
E_{f\text{-OPM1}}&=&-2J_1 - 3J_2 +6J_3,\label{eqn:3}\\
E_{f\text{-OPM2}}&=&2J_1 - 3J_2 -6J_3,\label{eqn:4}\\
E_{\text{FM}}&=&2J_1 + 6 J_2 + 12 J_3,\label{eqn:5}\\
E_{\text{Stripe-2}/\text{Zigzag-2}}&=&2J_1 - 2 J_2 - 4 J_3.\label{eqn:6}
\end{eqnarray}
For the parity-mixing state of $f\text{-OPM1}+g\text{-AM1}$, the energy is $-2J_1 - (3J_2 -6J_3)(\cos^2\theta-2\sin^2\theta)$, where $\theta$ is the canted angle between the magnetic moment and $ab$-plane. It cannot be ground state under this classical and isotropic approximation since $\cos^2\theta-2\sin^2\theta$ is a monotonic decreasing function when $\theta\in[0,\pi/2]$ (either $f$-OPM1 or $g$-AM1 as ground state). Quantum fluctuations or anisotropic exchange couplings may induce a certain degree of mixing~\cite{Triangular_Lattice}, as confirmed by our DFT calculation.

We plot a $J_2/|J_1|-J_3/|J_1|$ phase diagram for the magnetic ground state of this model in Fig.~\ref{fig:phase}(g) and (h) for AFM and FM $J_1$ coupling, respectively. When $J_3=0$, the magnetic ground state is independently determined by the interlayer and intralayer NN couplings, so there are four different ground states: (i) $f$-OPM1 when $J_1>0$ and $J_2>0$; (ii) $g$-AM1 when $J_1>0$ and $J_2<0$; (iii) $f$-OPM2 when $J_1<0$ and $J_2>0$; (iv) FM when $J_1<0$ and $J_2<0$. We find that finite $J_3$ plays important roles in determining the phase diagram although $J_3$ is smaller than $J_1$. This is because the multiplicity of the $J_3$ couplings are larger than that of $J_1$ and $J_2$. For instance, a FM $J_3$ coupling could lead to a nontrivial FM ground state even $J_1$ or $J_2$ is AFM coupling. Meanwhile, a FM $J_3$ coupling also tends to stabilize the $f$-OPM1 state even $J_2$ or $J_1$ is FM coupling. This will result in strong competition between FM and $f$-OPM1 state, as shown in the bottom-left of Fig.~\ref{fig:phase}(g). While, an AFM $J_3$ coupling will stabilize the $g$-AM1 state even $J_2$ is AFM coupling or $J_1$ is FM coupling, and it will also stabilize the $f$-OPM2 state even $J_1$ is AFM coupling or $J_2$ is FM coupling. 

We note that, in this classical $J_1$-$J_2$-$J_3$ model, the $g$-AM2, stripe and zigzag states can only become ground state at the phase boundary $2J_3=J_2$ ($J_1>0$) or $2J_3=-J_2$ ($J_1<0$), but they are still degenerate with $g$-AM1 and $f$-OPM1 at $2J_3=J_2$ [Fig.~\ref{fig:phase}(g)], or degenerate with $f$-OPM2 and FM at $2J_3=-J_2$ [Fig.~\ref{fig:phase}(g)]. Several factors could drive them to be ground state. Firstly, since the effect of $J_2$ has been canceled out by $J_3$, longer intralayer exchange couplings start to play roles. For example, if the intralayer NNN coupling $J_4$ is considered, Stripe-1 and Zigzag-1 states may become ground state when $-J_4<J_2-2J_3<8J_4$ and $-\frac{2}{3}J_4<J_2-2J_3<\frac{16}{3}J_4$, respectively, and the Stripe-2 and Zigzag-2 states may become ground states when $-J_4<J_2+2J_3<8J_4$ and $-\frac{2}{3}J_4<J_2+2J_3<\frac{16}{3}J_4$, respectively. Secondly, the symmetry of these AFM states are lower than that of FM, $g$-AM1, $f$-OPM1 and $f$-OPM2, so there will be concomitant lattice distortion that will further reduce energy. Thirdly, quantum fluctuations also play important roles in stabilizing these states. It is expected that the phase boundary of this classical model will be reshaped by these factors, so we roughly indicate these states near the phase boundary ($2J_3=J_2$ and $2J_3=-J_2$) by the dashed rectangles in the phase diagram. The existence of these states have been confirmed by our DFT calculations.

\textit{NiAs-type materials and phase transition.} We perform DFT calculations for several representative NiAs-type compounds including CrSb, MnTe, CrSe, CrTe, CrTe$_x$Se$_{1-x}$ and hexagonal FeTe, to verify the model and search more candidates of unconventional magnets. The DFT calculated magnetic exchange couplings $J_1$, $J_2$, $J_3$ and the total energies of different magnetic states are shown in Table~\ref{tab:energy}. The numbers in the parenthesis are the total energy calculated by Eqns.~\ref{eqn:1}-\ref{eqn:6} with the calculated $J_i$ as inputs. The ground state and the energy sequence of other magnetic states obtained by the $J_1$-$J_2$-$J_3$ model agree well with the DFT results. The only exception is FeTe where the ground state obtained by DFT and the model is slightly different, because it is very close to the phase boundary ($2J_3=-J_2$). 

The calculated $J_i$ is dominated by the interlayer NN coupling $J_1$. It is FM coupling for FeTe and AFM coupling for all other compounds. We have reproduced the $g$-AM1 ground state for CrSb and MnTe~\cite{CrSb_Neutron1,CrSb_Neutron2,CrSb_dft_gam1,CrSb_dft_gam2,MnTe_J4,MnTe_Neutron_dft}. We note that, although the calculated $J_2$ is AFM coupling for MnTe, the AFM $J_3$ coupling has overcome $J_2$ to stabilize a $g$-AM1 ground state, consistent with our model analysis and previous studies~\cite{MnTe_J,MnTe_J2,MnTe_J4}. We note that the important roles of $J_3$ in stabilizing the $g$-AM1 state in MnTe was also discussed by I. Mazin in Ref.~\cite{MnTe_J2}.

Our DFT calculations find that the ground state of CrSe is the mixed-parity umbrella-like state, but it is dominated by the $f$-OPM1 component. The calculated $J_3$ (-8.1 meV) is strong FM coupling, which helps to stabilize the $f$-OPM1 component. The calculated in-plane magnetic moment is about 3 $\mu_B$, and the out-of-plane moment is only about 0.1 $\mu_B$, which leads to a small canted angle $\theta\sim 2\degree$. This is consistent with our analysis based on the classical $J_1$-$J_2$-$J_3$ model. Since the $g$-AM1 component is much smaller than the $f$-OPM1 component, we still term the ground state of CrSe as $f$-OPM1 in the phase diagram [Fig.~\ref{fig:phase}(g)] and Table~\ref{tab:energy} for simplicity. We note that a neutron diffraction experiment performed on CrSe by L.M. Corliss \textit{et al.} in 1961 inferred that it has an umbrella-like NCL magnetic structure with large canted angle $\theta\sim45\degree$~\cite{CrSeneutron}. This discrepancy might be caused by sample quality. Studies have shown that there might be other coexisting phases such as Cr$_2$Se$_3$, Cr$_3$Se$_4$, Cr$_5$Se$_8$, Cr$_7$Se$_8$~\cite{CrSe_family2,Cr2Se3,CrSefamily,Cr3Se4,Cr0.68Se,CrSe2_Cr2Se3,CrSe_MBE,CrSefilms,CrSe_MBE2} when growing the CrSe samples. In particular, a recent study has inferred that there should be Cr$_6$Se$_8$ phase in the sample used in the neutron diffraction experiment~\cite{Cr6Se8}. This will significantly affect the results and analysis of neutron diffraction experiments. Further experiment with high quality CrSe sample is required to confirm the exact canted angle. 

\begin{figure}
    \centering
    \includegraphics[width=0.48\textwidth]{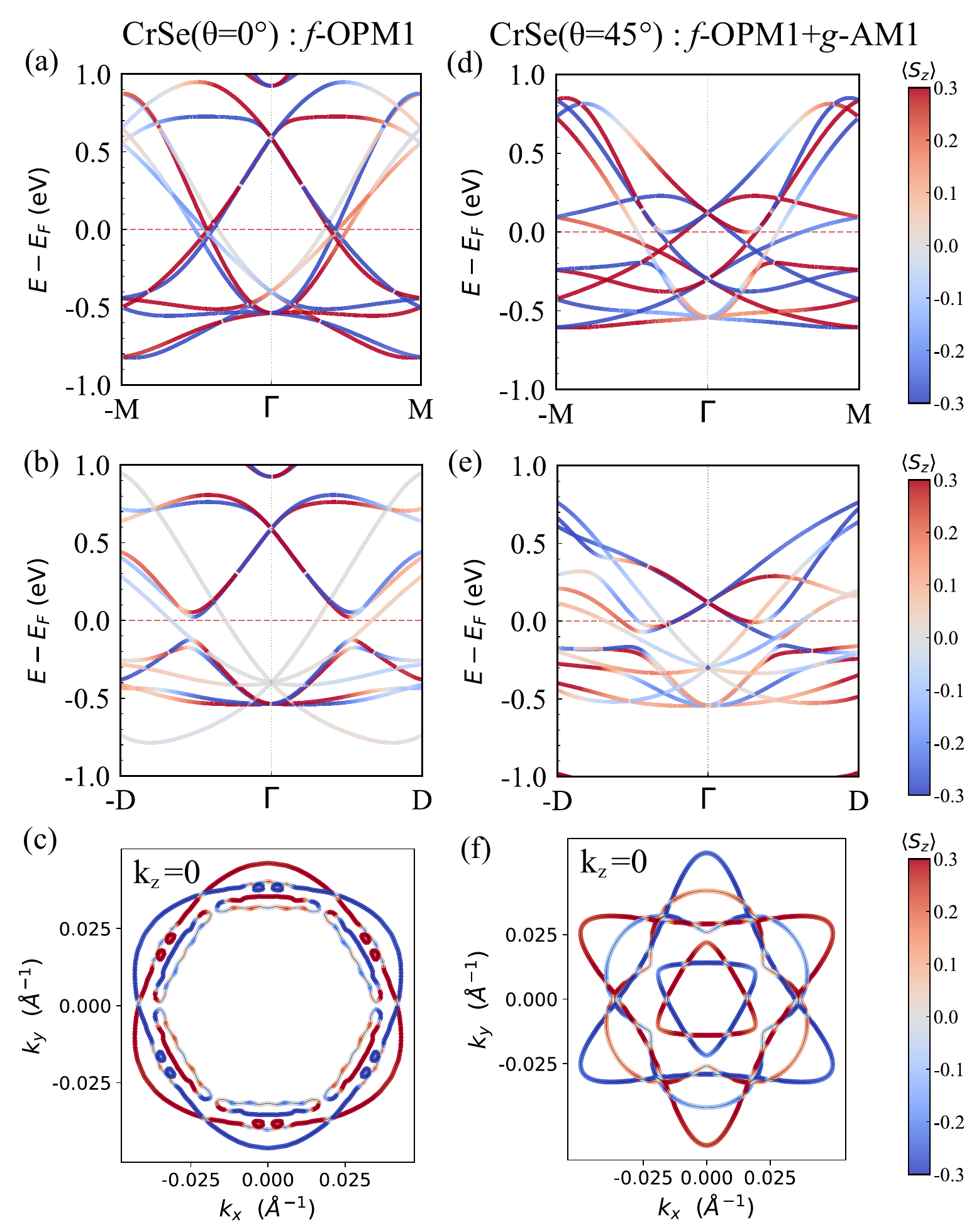}
    \caption{
     Band structures of CrSe along (a),(d) high-symmetry line $-M$-$\Gamma$-$M$ and (b),(e) low-symmetry line $-D$-$\Gamma$-$D$, where the fractional coordinate of $D$ is (0.3, 0.2, 0.2). (c),(f) Fermi surface at $k_z=0$ plane. (a)-(c) for pure $f$-OPM1 state ($\theta=0\degree$), and (d)-(f) for mixed-parity $f$-OPM1+$g$-AM1 state ($\theta=45\degree$). The color indicates the expectation value of $s_z(\mathbf{k})$. 
     }
    \label{fig:CrSe}
\end{figure}

To clearly show that the umbrella-like state is a mixed-parity of $s_z(\mathbf{k})$, we compare band structures and Fermi surfaces of CrSe between a pure $f$-OPM1 state ($\theta=0\degree$) and a hypothetical state with equally mixing of $f$-OPM1 and $g$-AM1 ($\theta=45\degree$) in Fig.~\ref{fig:CrSe}. Our calculations confirm $s_{xy}(\mathbf{k})=0$ and only $s_z(\mathbf{k})$ are nonzero. The pure $f$-OPM1 state shows odd-parity $f$-wave spin splitting in both high-symmetry ($k_z=0$ plane, $-M$-$\Gamma$-$M$ line) and low-symmetry ($-D$-$\Gamma$-$D$) plane/lines [Figs.~\ref{fig:CrSe}(a)-(c)]. The mixed-parity state exhibits only odd-parity $f$-wave spin splitting induced by the $f$-OPM1 component in the $k_z=0$ plane [Figs.~\ref{fig:CrSe}(d),(f)], since the $g$-AM1 component does not induce spin splitting in this plane. While, the mixed-parity state induces both even-parity [$s_z(\mathbf{k})=s_z(-\mathbf{k})$] and odd-parity [$s_z(\mathbf{k})=-s_z(-\mathbf{k})$] spin splitting along the low-symmetry line $-D$-$\Gamma$-$D$, so the net result is $s_z(\mathbf{k})\ne-s_z(-\mathbf{k})$, as shown in Fig.~\ref{fig:CrSe}(e).

We apply in-plane biaxial strain (changing lattice parameter $a$, with fixed $c$) on CrSe to explore possible phase transitions [Fig.~\ref{fig:phase}(i)]. The calculated $J_1$ and $J_2$ stay AFM coupling, while $J_3$ gradually changes from FM coupling to AFM coupling as decreasing $a$. This drives phase transitions from $f$-OPM1 to $g$-AM2 (0.97$a_0$), and then to Zigzag-1 (0.95$a_0$), and finally to $g$-AM1, because AFM $J_3$ coupling tends to stabilize the $g$-AM1 state. We note that $2J_3\approx J_2$ in the Zigzag-1 state. This confirms the existence of Zigzag-1 and $g$-AM2 state near the phase boundary between $g$-AM1 and $f$-OPM1, consistent with our analysis based on the $J_1$-$J_2$-$J_3$ model.

\begin{figure}
    \centering
    \includegraphics[width=0.5\textwidth]{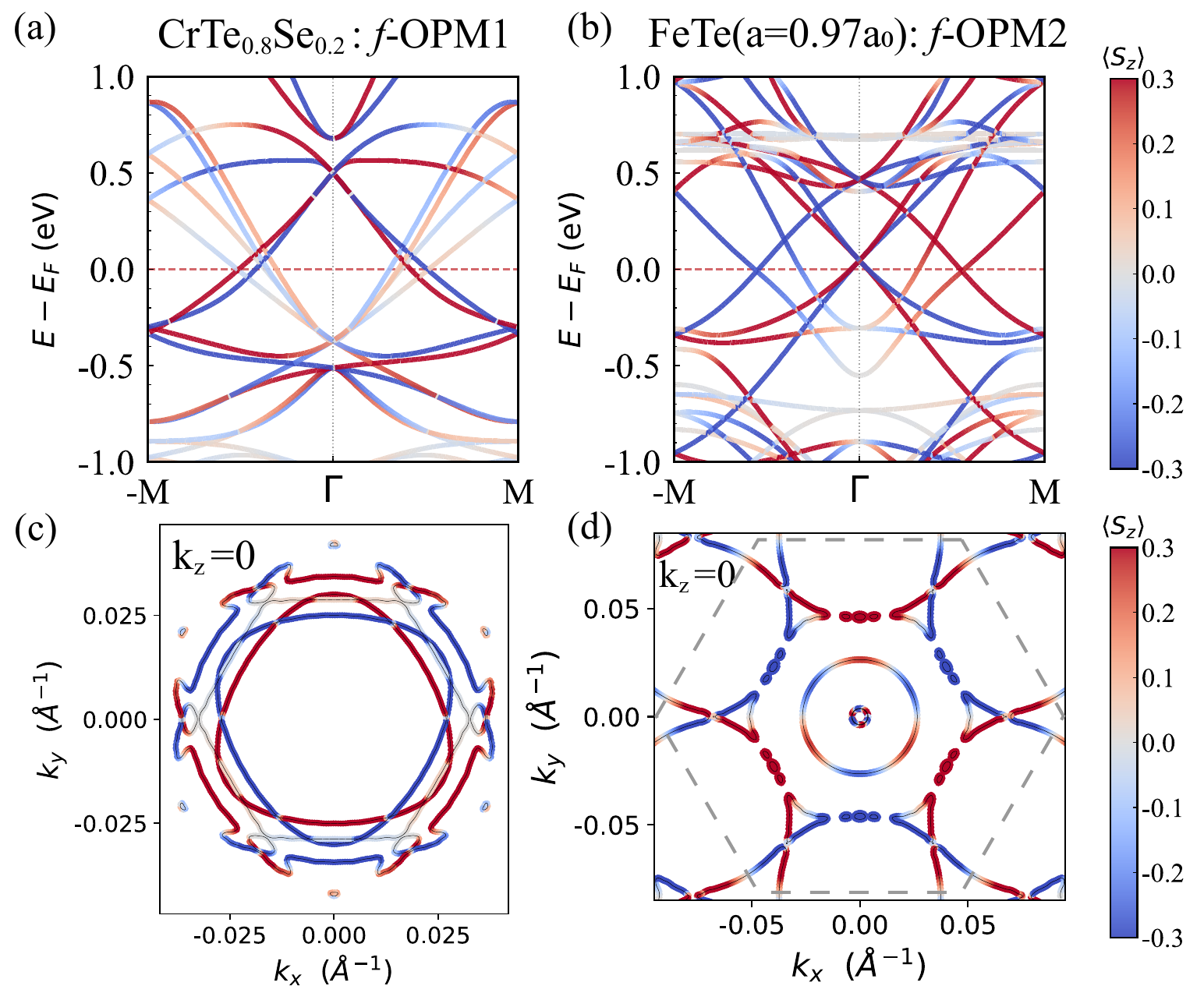}
    \caption{
     (a),(b) Band structures along $-M$-$\Gamma$-$M$ line. (c),(d) Fermi surface at $k_z=0$. (a),(c) for CrTe$_{0.8}$Se$_{0.2}$ and (b),(d) for FeTe with in-plane compressive strain ($a=0.97a_0$). The color indicates the expectation value of $s_z(\mathbf{k})$
     }
    \label{fig:FeTe}
\end{figure}

Our DFT calculations find a FM ground state for CrTe, which is consistent with experiments~\cite{Chiba:1960,HanJunbo:2020}. This FM state is nontrivial since the calculated $J_1$ (21 meV) is strong AFM coupling. It is still stabilized by the strong FM $J_3$ coupling (-13.2 meV) as we discussed above. Meanwhile, AFM $J_1$ and FM $J_3$ couplings also favor a $f$-OPM1 state. This leads to strong competition between the FM ground state and $f$-OPM1 state in CrTe, so it locates near the phase boundary as shown in the bottom-left of the phase diagram [Fig.~\ref{fig:phase}(g)]. The DFT calculated total energy of the $f$-OPM1 state is only 7 meV higher than that of the FM state (Table~\ref{tab:energy}). Therefore, the relative strength between $J_3$ and $J_1$ is the key factor determining either FM or $f$-OPM1 as ground state. Inspired by this, we replace some Te with Se to drive CrTe into the $f$-OPM1 state. This is confirmed by our DFT calculations shown in Fig.~\ref{fig:phase}(k), where the ground state of CrTe$_{1-x}$Se$_x$ becomes $f$-OPM1 when $x\ge0.2$. $|J_3/J_1|$ of CrTe$_{0.8}$Se$_{0.2}$ (0.55) becomes smaller than that of CrTe (0.63) [see Table~\ref{tab:energy} and Fig.~\ref{fig:phase}(g)]. The calculated in-plane magnetic moment is about 3.5 $\mu_B$. Similar to CrSe, there is also a very small canted moment ($\sim$ 0.1 $\mu_B$) along the $c$-axis. So, CrTe$_{1-x}$Se$_x$ is also a mixed-parity state with dominated $f$-OPM1 component. This is confirmed by the calculated band structure and Fermi surface at $k_z=0$ which exhibits a $f$-wave spin splitting of $s_z(\mathbf{k})$ [see Fig.~\ref{fig:FeTe}(a) and (c)]. 

We also apply in-plane biaxial strain to explore possible phase transition, which is shown in Fig.~\ref{fig:phase}(j). As decreasing $a$ (with fixed $c$), $J_1$ stays AFM coupling; $J_3$ tends to change from FM coupling to AFM coupling, while $J_2$ stays FM coupling and its strength first decreases and then significantly increases. This drives successive phase transitions from FM to $f$-OPM1 (0.97$a_0$), and then to $g$-AM2 (0.95$a_0$) and $g$-AM1 (0.92$a_0$). This demonstrates rich phase transitions among the multiple unconventional magnets in NiAs-type compounds.

For hexagonal FeTe, the calculated $J_1$ (-95.4 meV) and $J_3$ (-5.4 meV) are both FM coupling, and $J_2$ (12.3 meV) is AFM coupling. We note that $2J_3\approx -J_2$, so it is near the phase boundary between FM and $f$-OPM2. Our DFT calculations find Zigzag-2 as the ground state for FeTe, consistent with our analysis based on the $J_1$-$J_2$-$J_3$ model. We find pretty large distortions of the triangular lattice in this state, which further reduces energy such that it becomes ground state.  However, its energy is still close to that of $f$-OPM2 and FM states [see Table.~\ref{tab:energy}], indicating strong competition. Indeed, FM state was found in ultrathin film of hexagonal FeTe~\cite{Kang-FeTe:2020}.  We find that in-plane compressive strain could drive it to the $f$-OPM2 state [see Fig.~\ref{fig:phase}(l)]. For example, at $a=0.97a_0$, $f$-OPM2 becomes ground state, with 5 meV lower in energy than that of Zigzag-2 (Table~\ref{tab:energy}). The band structure and Fermi surface at $k_z=0$ show a $f$-wave spin splitting, as shown in Figs.~\ref{fig:FeTe}(b) and (d).

\textit{Conclusion and discussion.} To conclude, we have presented a global magnetic phase diagram for NiAs-type compounds, and we find multiple unconventional magnets including two $g$-wave AMs and two $f$-wave OPMs. A mixed-parity of $f$- and $g$-wave magnet could also naturally emerge in an umbrella-like NCL structure in this system. Our DFT calculations show that CrSe and CrTe$_{1-x}$Se$_x$ are generally in such mixed-parity state. The interlayer NNN coupling $J_3$ is found to be crucial in inducing strong competition between conventional and unconventional magnets. In particular, it could induce a nontrivial FM state even when $J_1$ is AFM coupling. This leads to  strong competition between FM and $f$-OPM1 states. It also induces competing states near the phase boundary setting by $2J_3=J_2$ and $2J_3=-J_2$. Inspired by this, we demonstrate that rich phase transitions and unconventional magnets could be realized by chemical doping or in-plane strain.
Our results demonstrate that the NiAs-type compounds is a good platform for realizing both even- and odd-parity as well as mixed-parity unconventional magnets, and the phase diagram provides a guidance for design of such magnets. The predicted unconventional magnets CrSe, CrTe$_{1-x}$Se$_x$ and FeTe are metallic, which are beneficial to construct heterostructure with superconductors for realizing topological superconducting state via proximity effect.  

\textit{Note added.} We note that two recent works posted in arXiv server also discussed the mixed-parity magnets~\cite{Luo:2026,Zhuang:2026}. Ref.~\cite{Luo:2026} terms such states as unconstrained-parity magnet and candidate materials including CrSe are discussed. Ref.~\cite{Zhuang:2026} proposes using circularly polarized light to drive mixed-parity magnets in collinear magnetic orders.
 
\textit{Acknowledgement.} This project was supported by National Key R\&D Program of China (Grant No. 2023YFA1406304), the Quantum Science and Technology-National Science and Technology Major Project (No. 2021ZD0302803) and the New Cornerstone Science Foundation. The calculations were performed in Hefei Advanced Computing Center, China.

%\pagebreak
\bibliography{main}

\end{document}

\title{Supplementary Materials for ``Global magnetic phase diagram and multiple unconventional magnets in NiAs-type compounds''}
\author{Shibo Shen}
\author{Yilin Wang}\email{yilinwang@ustc.edu.cn}
\affiliation{School of Emerging Technology, University of Science and Technology of China, Hefei 230026, China}
\affiliation{Hefei National Laboratory, University of Science and Technology of China, Hefei 230088, China}
\affiliation{New Cornerstone Science Laboratory, University of Science and Technology of China, Hefei, 230026, China}

\date{\today}
%\begin{abstract}
%This includes the computational details and the DFT results of FeSn.
%\end{abstract}

\maketitle

\section{\lowercase{\uppercase{C}omputational \uppercase{D}etails}}
Density functional theory (DFT) calculations are performed using the projector augmented wave (PAW)~\cite{paw} method as implemented in the Vienna \textit{Ab-initio} Simulation Package (VASP)~\cite{vasp}. The exchange-correlation functional of generalized gradient approximation (GGA-PBE) is used.  The cutoff of plane-wave energy is 500 eV. $\Gamma$-centered K-point grid of $16 \times 16 \times 8$, $10 \times 10 \times 8$, $8 \times 16 \times 8$ and $8 \times 6 \times 8$ are used for the $1\times1\times1$, $\sqrt{3}\times\sqrt{3}\times1$, $2\times1\times1$, and $2\times3\times1$ supercells, respectively. The Se-doping is simulated using the method of virtual crystal approximation (VCA)~\cite{vca}. The spin-orbital coupling is not considered in the calculations. The 2D Fermi surfaces are visualized and analyzed with Pyprocar~\cite{PyProcar1,PyProcar2}.

We use experimental lattice constants for CrSb, MnTe, and CrSe. For FeTe and CrTe, we perform full structural relaxation for all magnetic configurations. This is because these materials lie near the phase boundary, where the magnetic ground state is very sensitive to lattice constants and atomic positions. For structural relaxation, the criteria for forces convergence is less than 0.005 eV/\AA. The magnetic structures of Stripe-1/Stripe-2 and Zigzag-1/Zigzag-2 are shown in Fig.~\ref{fig:structures}.

The magnetic exchange coupligns $J_i$ are calculated based on the total energy difference method, with the help of the open-source OstravaJ code~\cite{oj}. To calculate $J_1$, $J_2$ and $J_3$, we employ a $2\sqrt{3}\times\sqrt{3}\times1$ supercell with a $\Gamma$-centered $5 \times 10 \times 8$ k-point grid. For CrSb, MnTe, CrSe, the experimental crystal structures are used. For CrTe, CrTe$_{0.8}$Se$_{0.2}$ and FeTe, the relaxed crystal structures of the ground states are used. The lattice constants used for calculating $J_i$ and the Cr-Cr distances for $J_1$, $J_2$, $J_3$ are listed in Table.~\ref{tab:lattice_constants}. We use the convention that the spin vectors $\vec{S}_i$ are normalized to 1, so that the parameters $J_i$ are in units of energy.

\section{\lowercase{\uppercase{T}otal \uppercase{E}nergies \uppercase{F}ormula  \uppercase{C}onsidering the \uppercase{I}ntralayer \uppercase{NNN} \uppercase{C}oupling \uppercase{$J_4$} }}
\begin{figure}[htbp]
\centering
\includegraphics[width=1\textwidth]{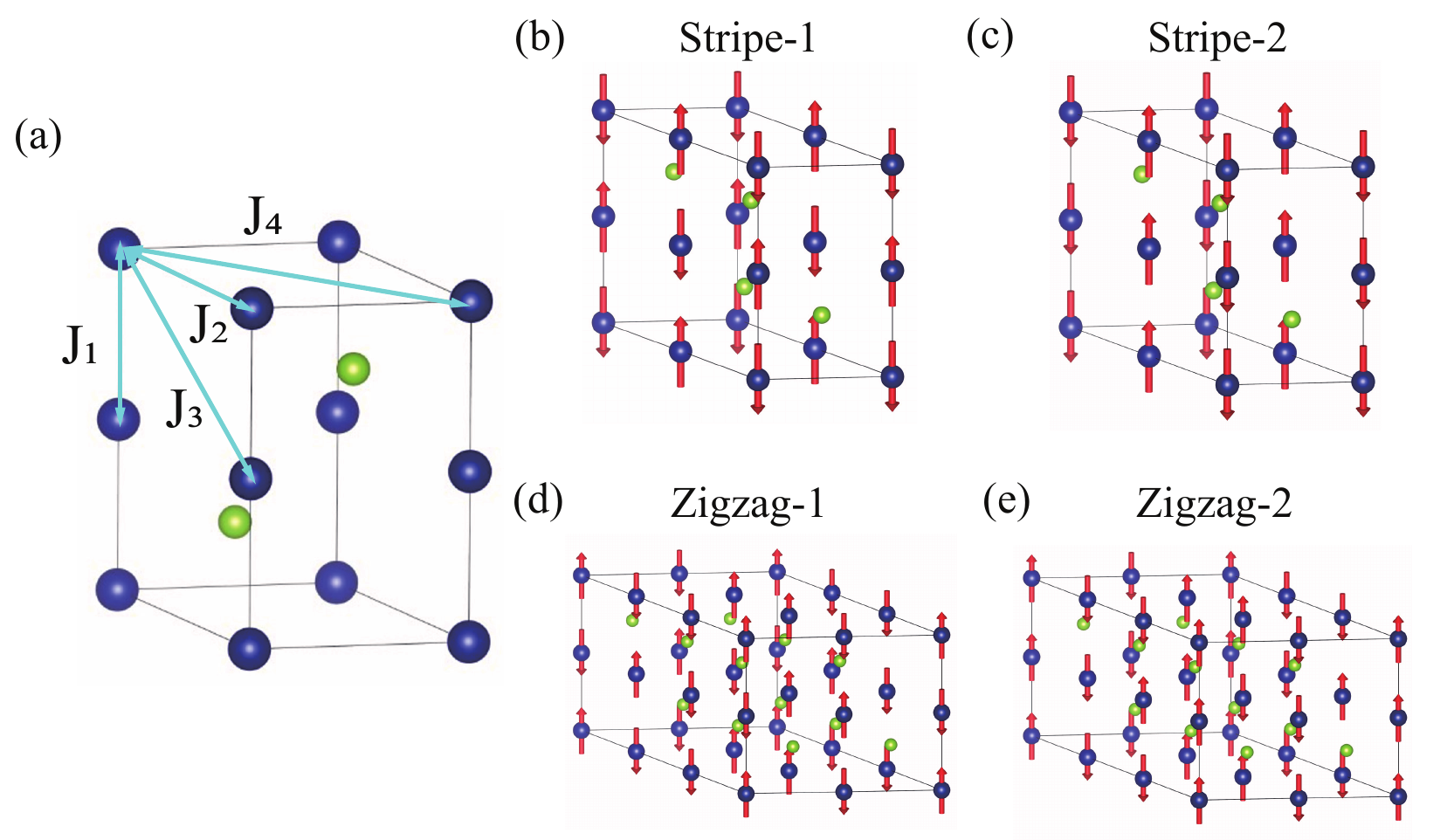}
\caption{(a) Definition of the exchange interactions $J_1$-$J_4$ for the NiAs-type structure. (b)-(e) Magnetic structures of four conventional AFM orders.}
\label{fig:structures}
\end{figure}

As shown in Fig.~\ref{fig:structures}(a), we consider the intralayer NNN coupling $J_4$. The formula of total energy are,
\begin{eqnarray}
E_{g\text{-AM1}}&=&-2J_1 + 6J_2 -12J_3+6J_4,\label{eqn:1}\\
E_{g\text{-AM2}}&=&-2J_1 - 2J_2 +4J_3+6J_4,\label{eqn:2}\\
E_{\text{Stripe-1}}&=&-2J_1 - 2J_2 +4J_3-2J_4,\label{eqn:3}\\
E_{\text{Zigzag-1}}&=&-2J_1 - 2J_2 +4J_3+\frac{2}{3}J_4,\label{eqn:4}\\
E_{f\text{-OPM1}}&=&-2J_1 - 3J_2 +6J_3+6J_4,\label{eqn:5}\\
E_{f\text{-OPM2}}&=&2J_1 - 3J_2 -6J_3+6J_4.\label{eqn:6}\\
E_{\text{FM}}&=&2J_1 + 6 J_2 + 12 J_3+6J_4,\label{eqn:7}\\
E_{\text{Stripe-2}}&=&2J_1 - 2 J_2 - 4 J_3-2J_4.\label{eqn:8}\\
E_{\text{Zigzag-2}}&=&2J_1 - 2 J_2 - 4 J_3+\frac{2}{3}J_4.\label{eqn:9}
\end{eqnarray}

Based on Eqs.~\ref{eqn:1}-\ref{eqn:9}, we obtain the  critical condition  for the Stripe-1, Zigzag-1, Stripe-2, and Zigzag-2 states to become ground states as follows,

\begin{eqnarray}
\text{Stripe-1}&:& J_1>0 \quad \text{and} \quad -J_4<J_2-2J_3<8J_4,\label{eqn:10}\\
\text{Zigzag-1}&:& J_1>0 \quad \text{and} \quad -\frac{2}{3}J_4<J_2-2J_3<\frac{16}{3}J_4,\label{eqn:11}\\
\text{Stripe-2}&:& J_1<0 \quad \text{and} \quad -J_4<J_2+2J_3<8J_4,\label{eqn:12}\\
\text{Zigzag-2}&:& J_1<0 \quad \text{and} \quad -\frac{2}{3}J_4<J_2+2J_3<\frac{16}{3}J_4.\label{eqn:13}
\end{eqnarray}

\begin{table}[h]
\caption{The lattice constants and the Cr-Cr distances for the exchange couplings $J_1$ to $J_4$ of NiAs-type materials.}
\label{tab:parameters}
\begin{ruledtabular}
\begin{tabular}{cccccccccc}
System & a (\AA) & c (\AA) & $J_1$ (\AA) & $J_2$ (\AA) & $J_3$ (\AA) & $J_4$ (\AA) \\ \hline
CrSb\cite{crsb} & 4.108 & 5.440 & 2.720 & 4.108 & 4.927 & 7.115\\
MnTe\cite{mnte} & 4.110 & 6.700 & 3.350 & 4.110 & 5.302 & 7.119\\
CrSe\cite{CrSe} & 3.684 & 6.019  & 3.010 & 3.684 & 4.757 & 6.381\\
CrTe \textsuperscript{a}  & 4.129 & 6.295 & 3.147 & 4.129 & 5.192 & 7.152  \\
CrTe$_{0.8}$Se$_{0.2}$ \textsuperscript{a}  & 4.068 & 6.216 & 3.108 & 4.068 & 5.120 & 7.046\\
FeTe \textsuperscript{a}  & 4.142 & 5.092 & 2.546 & 4.142 & 4.862 & 7.175  \\
FeTe(a=0.97a$_0$) \textsuperscript{a}  & 4.018 & 5.092 & 2.546 & 4.018 & 4.757 & 6.960\\
\end{tabular}
\end{ruledtabular}
    \label{tab:lattice_constants}
\footnotetext{Lattice constants obtained from DFT relaxation.}
\end{table}

\begin{figure}[htbp]
\centering
\includegraphics[width=1\textwidth]{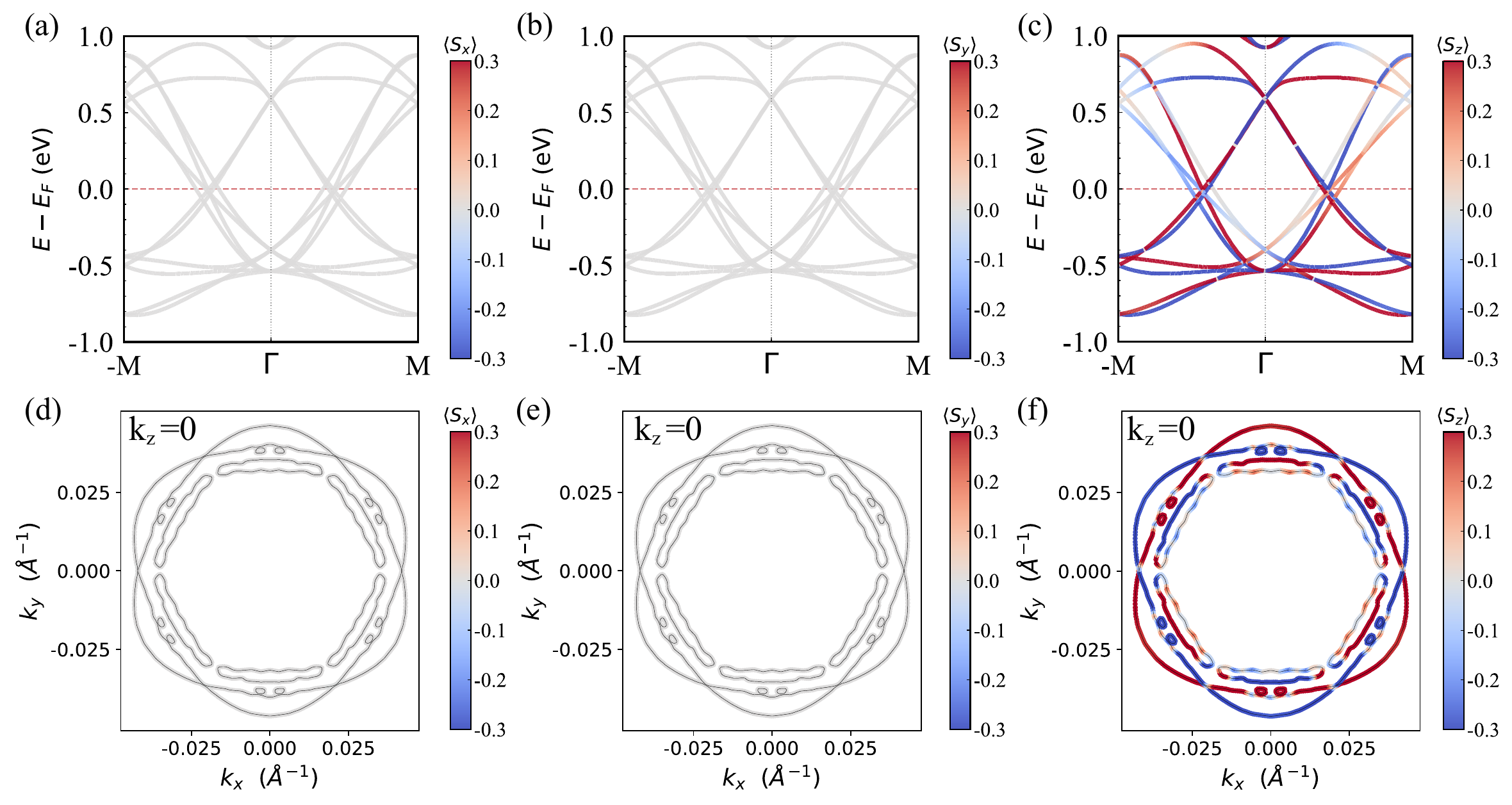}
\caption{Band structures and Fermi surface of CrSe in the $f$-OPM1 state. It demonstrates that $s_z(\mathbf{k})=-s_z(\mathbf{-k}), s_{x,y}\mathbf(\mathbf{k})=0$.}
\label{fig:CrSe_band}
\end{figure}

\begin{figure}[htbp]
\centering
\includegraphics[width=0.8\textwidth]{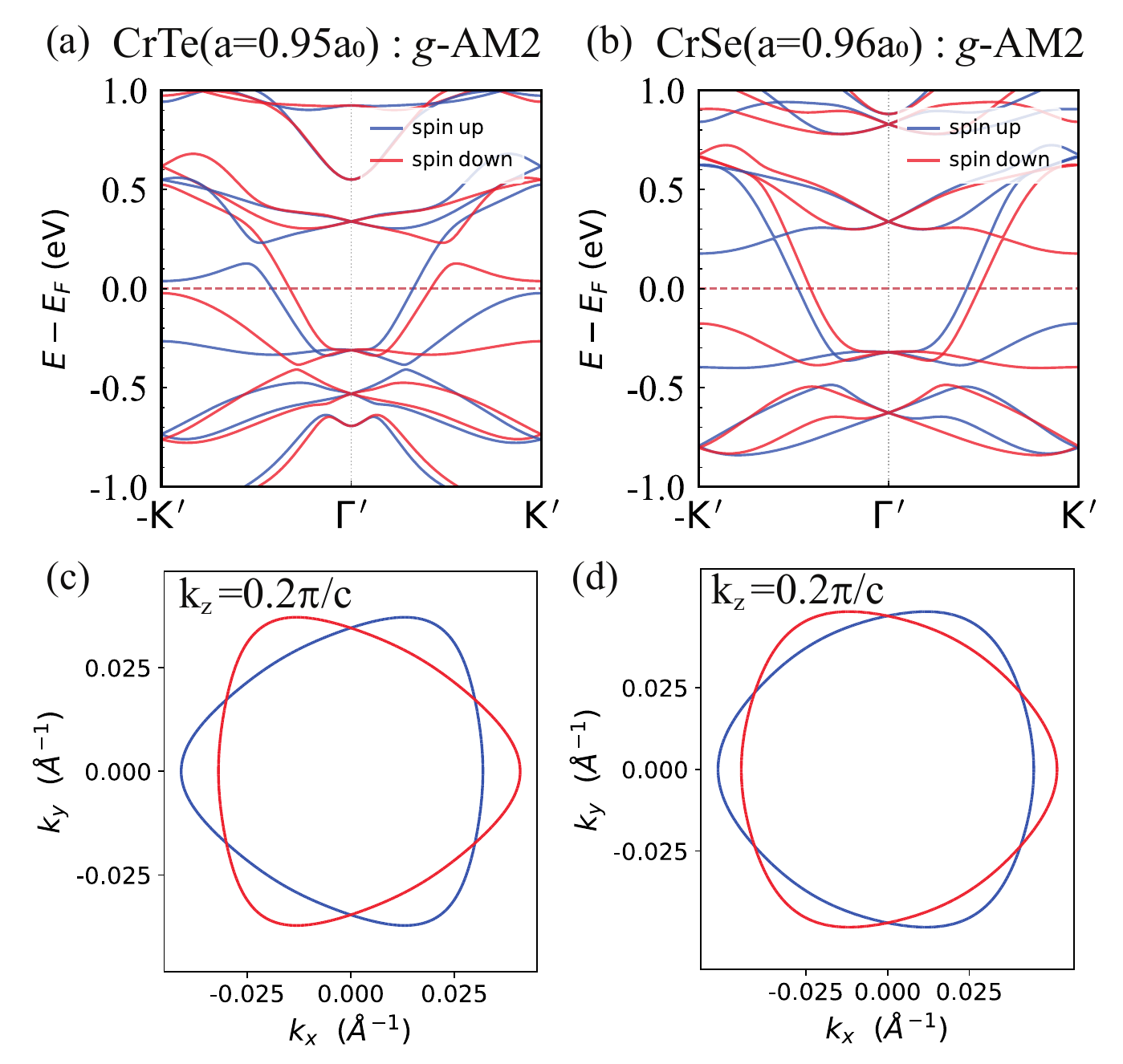}
\caption{Band structure and Fermi surface of CrTe (a=0.95a$_0$) and CrSe (a=0.96$a_0$) in the $g$-AM2 phase: (a),(b) Band structure along $-K'$-$\Gamma'$-$K'$. (c),(d) Fermi surface at $k_z=0.2\pi/c$. Blue and red lines correspond to spin-up and spin-down states.}
\label{fig:CrTe(0.95a)}
\end{figure}

\begin{figure}[htbp]
\centering
\includegraphics[width=0.75\textwidth]{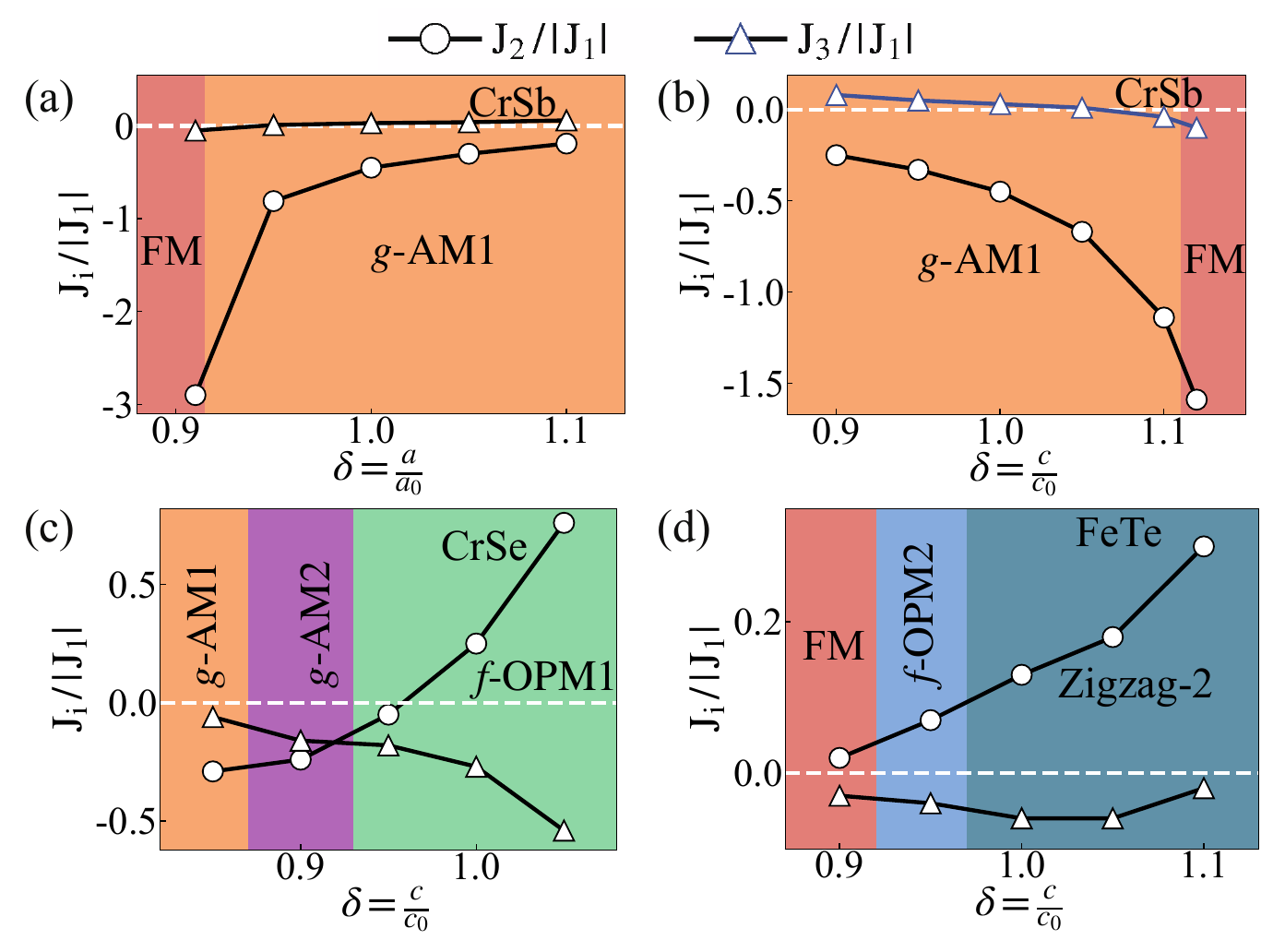}
\caption{DFT calculated $J_2/|J_1|$, $J_3/|J_1|$ and phase transitions of CrSb, CrSe and FeTe as changing lattice parameters $a$ or $c$.}
\label{fig:change_z}
\end{figure}

\pagebreak
\bibliography{suppl}